\documentclass[preprintnumbers,amsmath,amssymb,showpacs,twocolumn]{revtex4}

\usepackage{graphicx}
\usepackage{dcolumn}
\usepackage{bm}

\begin{document}

\title{ A  method for teleporting an unknown quantum  state and its application}

\author{ Feng-Li Yan $^{1,2}$, Hai-Rui Huo $^{1}$}

\affiliation {
$^1$  College of Physics Sciences and Information Engineering, Hebei Normal University, Shijiazhuang 050016, China\\
$^2$ CCAST (World Laboratory), P.O. Box 8730, Beijing 100080,
China }

\date{\today}

\begin{abstract}
 We suggest a  method for teleporting an unknown quantum state.  In this method  the sender Alice first
 uses a Controlled-Not operation  on the particle in  the unknown quantum
 state and an ancillary  particle which she wants to send to the receiver Bob. Then she sends ancillary  particle to Bob.
 When Alice is informed by Bob that the ancillary  particle is received, she performs a local measurement on the
 particle and sends Bob the outcome of the local measurement via a classical
 channel. Depending on the outcome Bob can restore the unknown quantum
  state, which Alice destroyed, on the ancillary particle successfully.
  As an application of this method we propose a quantum secure direct communication protocol.
  \end{abstract}


\maketitle

One of the most fascinating subjects in fundamental quantum theory
is  quantum information. It presents a new perspective for all
foundational and interpretational issues and highlights new
essential differences between classical and quantum theory.
Quantum entanglement has an extremely important position  in the
 processing and transmitting of quantum information. Many useful applications such as
quantum dense coding \cite {s1}, certain types of quantum key
distribution \cite {s2} and Greenberger-Horne-Zeilinger
correlations \cite {s3} are based on the existence of quantum
entanglement.

One of the extremely striking exhibitions of entanglement is quantum teleportation. In 1993, Bennett et al
suggested a quantum method of teleportation \cite {s4}, by which an unknown quantum state of one qubit can be
transmitted from one place to another with the aid of some  classical communication, provided that the sender
Alice and the receiver  Bob have previously shared halves of a two qubit entangled state.  At present,
teleportation has been generalized to many cases \cite {s5,s6,s7, s8,s9,s10,s11,s12} in theoretical aspect.
Experimentally the teleportation of a photon polarization state has been demonstrated \cite {s13, s14}. In
virtue of a two-mode squeezed vacuum state, teleportation of a coherent state corresponding to continuous
variable system was realized also in the laboratory \cite {s15}.

Another important part of quantum information is quantum cryptology. Amazingly, the principles of quantum
mechanics such as the uncertainty principle and quantum correlations have now provided the foundation stone to a
new approach to cryptography - quantum cryptography. It has been believed that quantum cryptography can solve
many problems that are impossible from the perspective of conventional cryptography. Since the first quantum
cryptography  protocol  using quantum mechanics to distribute keys was proposed by Bennett and Brassard in 1984
\cite {s16}, numerous quantum cryptographic protocols has been proposed \cite {s17, s18, s19, s20, s21, s22,
s23, s24,s25,s26,s27,s28,s29,s30}. Recently, Shimizu and Imoto \cite {s31,s32} and Beige et al \cite {s33}
proposed novel quantum secure direct communication (QSDC) schemes. In these protocols the sender Alice and the
receiver Bob communicate messages directly without a shared secret key to encrypt them and the message is
deterministically sent through the quantum channel, but can be read only after obtaining additional classical
information for each qubit. After that a few theoretical schemes for QSDC were put forward \cite
{s34,s35,s36,s37}. However,
 in all these QSDC schemes one must send the qubits with secret
messages in a public channel. Therefore, a potential eavesdropper, Eve, can attack the qubits with secret
messages in transmission. In order to conquer this limitation several QSDC protocols with using quantum
correlations of Einstein-Podolsky-Rosen (EPR) pairs or Greenberger-Horne-Zeilinger (GHZ) states and
teleportation, have been proposed  \cite {s38,s39,s40,s41}. In the framework of these protocols there is not a
transmission of qubits carrying the secret messages in a public channel, so that  the secret messages can not be
attacked in transmission.

This letter contains twofold purposes. One is to present a method
for teleporting an unknown quantum state; the other is
 to  provide  a QSDC protocol by applying the  method for teleporting.

We describe the process for teleporting an unknown quantum state
as follows. Suppose that an unknown quantum state of particle 1 is
\begin{equation}
|\phi\rangle_1=\alpha|0\rangle_1+\beta|1\rangle_1,
\end{equation}
with $|\alpha|^2+|\beta|^2=1$, and particle 1 is in Alice's
possession. Alice wants to teleport the state $|\phi\rangle$ to
the receiver Bob. To do so Alice introduces an ancillary particle
2 initially in a known state $|0\rangle_2$. Then she  makes a
Controlled-Not operation on two particles with particle 1 and
particle 2 being  controlled bit  and  target bit respectively. By
completing this unitary operation, the quantum state of the whole
system consisting of two particles 1 and 2 becomes
\begin{equation}
|\Psi\rangle_{12}=\alpha|00\rangle_{12}+\beta|11\rangle_{12}.
\end{equation}
Then Alice sends the particle 2 to Bob and she will be informed
via a classical channel when Bob receives the particle 2. A
algebraic rearrangement of Eq.(2) in terms of $|+\rangle_1=\frac
{1}{\sqrt 2}(|0\rangle+|1\rangle)_1$ and $|-\rangle_1=\frac
{1}{\sqrt 2}(|0\rangle-|1\rangle)_1$ leads to
\begin{eqnarray}
|\Psi\rangle_{12}
 &&=
\frac {1}{\sqrt
2}|+\rangle_1(\alpha|0\rangle+\beta|1\rangle)_{2}\nonumber\\
&&~~+\frac {1}{\sqrt
2}|-\rangle_1(\alpha|0\rangle-\beta|1\rangle)_{2}.
\end{eqnarray}
Evidently, if  Alice performs a local measurement on her particle
1 in the basis $\{ |+\rangle_1, |-\rangle_1\}$, then regardless of
the identity of $|\phi\rangle_1$, each outcome will occur with
equal probability $\frac {1}{2}$. Therefore,  this measurement can
not give Alice  information at all about the identity of the state
and it would collapse  the resulting state of  particle 2 to
\begin{eqnarray}
&&(\alpha|0\rangle+\beta|1\rangle)_{2}=I|\phi\rangle_2\equiv U_0|\phi\rangle_2,\nonumber\\
&&(\alpha|0\rangle-\beta|1\rangle)_{2}=\sigma_z|\phi\rangle_2\equiv
U_1|\phi\rangle_2
\end{eqnarray}
respectively.  Here $I$ is the identity operator, and $\sigma_z$
indicates Pauli operator.  Obviously, in each case the state of
 particle 2 is related to $|\phi\rangle_2$ by a fixed unitary
transformation $U_i$ $(i=0,1)$ independent of the identity of
$|\phi\rangle$. Hence if Alice tells  Bob  her actual measurement
outcome, then Bob will be capable of  applying the corresponding
inverse transformation $U_i^{-1}$ to his particle 2, restoring it
to state $|\phi\rangle$ in every case, i.e. Bob can convert the
state of particle 2 into an exact replica of the unknown quantum
state which Alice destroyed.

 Hence   Alice is able to
 communicate to Bob the full quantum information of
 $|\phi\rangle$ by this method.
 As a result of the process, Alice learns nothing whatever about identity of
 $|\phi\rangle$.  In the process of teleporting an unknown quantum state, Bob is
 left with a perfect instance of
 $|\phi\rangle$ and hence no participants can gain any further information about its
 identity.

 As a matter of fact,   in comparing our protocol with the standard teleportation
 scheme \cite {s4} our protocol is more economical one.  On one hand,
 in present scheme  only  two particles are involved, that is not the same as that of standard teleportation
 scheme, in which  three particles  take part in. On the other hand  the classical
 information needed to be transmitted is one bit,
 more economical than that of the standard one in
 which two bits classical information must be sent.

 Apparently, this method can be generalized to multiple-particle
 cases easily.

 As an application of this method for teleporting an unknown quantum state, now we would like to suggest
 a
 QSDC protocol, which will be stated  as follows.

  Assume that the sender Alice wants to communicate important messages to  the receiver Bob,
  and they share  a set of EPR pairs, the
maximally entangled pair  in the Bell state
\begin{equation}
|\Phi^+\rangle_{AB}=\frac {1}{\sqrt
2}(|00\rangle_{AB}+|11\rangle_{AB}).
\end{equation}
Obviously, one approach to prepare the EPR pair can be that  Alice
first  makes one particle in the state $\frac {1}{\sqrt
2}(|0\rangle+|1\rangle)$, and performs the Controlled-Not
operation on two particles, where  the state of other particle is
 $|0\rangle$ initially.  It leads to that the pair of the
two particles is in the Bell state stated in Eq.(5). Then Alice sends half of the pair to Bob. They can also
obtain EPR pairs in other method as mentioned in Ref. \cite {s38}. Before communication they must do some tests,
for example to use the schemes testing the security of EPR pairs (quantum channel) in Refs. \cite {s2, s18, s26,
s36, s38}. Passing the test asserts that they continue to hold sufficiently pure, entangled quantum states.
However, if tampering has occurred, these EPR pairs should be discarded and new EPR pairs should be constructed
again.

After insuring the security of quantum channel, Alice and Bob begin to  communicate.   Alice performs operation
$I$ or $\sigma_z$ on the particle A according to the secret messages. For instance  if the secrete messages are
00101011, then the operation Alice performed should be $II\sigma_zI\sigma_z I\sigma_z\sigma_z$ on eight EPR
pairs respectively. In other words if the secret message is 0 then the EPR pair is unchanged; otherwise  EPR
pair state should be $\frac {1}{\sqrt 2}(|00\rangle_{AB} -|11\rangle_{AB})$. Thus Alice completes the
transformation between the secret messages and the states of EPR pairs, i.e. 0 and 1 of the secret messages
correspond to $ |\Phi^+\rangle_{AB}=\frac {1}{\sqrt 2}(|00\rangle_{AB}+|11\rangle_{AB})$ and  $
|\Phi^-\rangle_{AB}=\frac {1}{\sqrt 2}(|00\rangle_{AB}-|11\rangle_{AB})$ respectively. After that Alice can use
the method for teleporting an unknown quantum state stated above to transfer the secret message to Bob. Alice
makes firstly  the local measurement on her particle A in the basis $\{|+\rangle_A, |-\rangle_A\}$ and tells Bob
the outcome of local measurement via a classical channel. According to the measurement results Bob makes the
unitary transformation on particle B, provided that if Alice's the measurement result is $|+\rangle$, then Bob
do not apply any operator to his particle B, otherwise
 Bob performs $\sigma_z$ operation to
his particle B. Thus the secret messages have been  transferred to
the states of the particles in Bob's possession. Finally Bob makes
a local measurement on the particle B in the basis $\{|+\rangle_B,
|-\rangle_B\}$ which will tell him the secret message accurately.

This QSDC protocol has two notable features. One is  that in our scheme the classical information resource is
saved on since one bit classical information  is only wanted to transmit one bit secret message in comparing
with the protocol in Ref.\cite {s38} in which in order to transfer one bit secret message two bit classical
messages must be sent via a classical channel. Another is that in present  protocol there is not a particle with
the secret message transmitting between Alice and Bob, hence Eve can not attack the message quantum bit in
transmission.

Evidently, if the quantum
 channel is perfect EPR pairs, the protocol must be a safety one.
 As mentioned in Ref. \cite {s38}, by using the schemes testing the security of
 EPR pairs,  a perfect quantum channel can be obtained. So Alice
 and Bob can communicate the secret messages by this protocol safely.

In summary  we give a  method for teleporting an unknown quantum
state, in which  the sender Alice first
 uses a Controlled-Not operation  on the particle in  the unknown quantum
 state and an ancillary  particle.
 Then the ancillary  particle is sent to Bob.
 When Alice knows that the ancillary  particle is received by Bob,
 she performs a local measurement on the
 particle and sends Bob the outcome of the local measurement via a classical
 channel. According to  the result of the measurement Bob can restore the unknown quantum
  state, which Alice destroyed, on the ancillary particle.
  By the way  a QSDC communication protocol is proposed.
   The communication is based on EPR pairs functioning as quantum
channel. The QSDC protocol not only saves the classical information, but also    defends signal against
interference.

Apparently, in our QSDC scheme we requires the technique of quantum storage to guarantee the EPR pairs being in
the maximally entangled state. As a matter of fact this technique is not fully developed at present. However it
is of vital importance to quantum information, and there has been great interest in developing it \cite {s42,
s43, s44}. We believe that it
 will be available in the future, and
hope our scheme will be realized in  experiment.

 \acknowledgments This work was supported by Hebei Natural Science Foundation of China under Grant No:
A2004000141 and No: A2005000140, and  Natural Science Foundation of Hebei Normal University.


\begin{thebibliography}{s2}
\bibitem{s1} C. H. Bennett and S. J. Wiesner, Phys. Rev. Lett.  {\bf 69}, 2881 (1992).
\bibitem{s2} A. K. Ekert, Phys. Rev. Lett. {\bf 67},  661 (1991).
\bibitem{s3} D. M. Greenberger, M. A. Horne, A. Shimony, and A.
Zeilinger, Am. J. Phys. {\bf 58}, 1131 (1990).
\bibitem{s4} C. H. Bennett, G. Brassard, C. Crepeau, R. Jozsa,
A. Peres, and W. K. Wootters,  Phys. Rev. Lett. {\bf 70}, 1895  (1993).
\bibitem{s5} M. Ikram, S. Y. Zhu, and M. S.  Zubairy, Phys. Rev. A{\bf 62}, 022307 (2000).
\bibitem{s6} W. L. Li, C. F. Li, and G. C. Guo, Phys. Rev. A{\bf 61},  034301 (2000).
\bibitem{s7} V. N. Gorbachev  and A. I. Trubilko, J. Exp. Theor. Phys.  {\bf 91}, 894 (2000).
\bibitem{s8} H. Lu and  G. C. Guo, Phys. Lett.   A{\bf 276}, 209 (2000).
\bibitem{s9} J. Lee and M. S.  Kim, Phys. Rev. Lett. {\bf 84},  4236 (2000).
\bibitem{s10} B. S. Shi, Y. K. Jiang, and G.C. Guo,  Phys. Lett. A{\bf 268}, 161 (2000).
\bibitem{s11} J. D. Zhou, G. Hou, and Y. D. Zhang, Phys. Rev. A {\bf 64}, 012301 (2001).
\bibitem{s12} F. L. Yan and D. Wang, Phys. Lett. A {\bf 316}, 297 (2003).
\bibitem{s13} D. Bouwmeester, J. W. Pan, K. Mattle, M. Eible, H. Weinfurter, and A. Zeilinger,  Nature  {\bf 390}, 575 (1997).
\bibitem{s14} D. Boschi,  S. Branca, F. DeMartini, L. Harely, and S. Popescu, Phys. Rev. Lett.   {\bf 80}, 1121 (1998).
\bibitem{s15} A. Furusawa, J. L. Sorensen, S. L. Brawnstein, C. A. Fuchs, H. J. Kimble, and E. Polizk,  Science  {\bf 282}, 706 (1998).
\bibitem{s16} C. H. Bennett and G. Brassard,  Proc. IEEE Int. Conf. on
Computers, Systems and Signal Processing, Bangalore, India, (IEEE, New York, 1984),  pp. 175-179.
\bibitem{s17} C. H. Bennett, Phys. Rev. Lett.  {\bf 68},   3121 (1992).
\bibitem{s18} C. H. Bennett, G. Brassard, and N. D. Mermin,  Phys. Rev. Lett. {\bf 68}, 557 (1992).
\bibitem{s19} L. Goldenberg and L. Vaidman,  Phys. Rev. Lett.  {\bf 75}, 1239 (1995).
\bibitem{s20} B. Huttner, N. Imoto, N. Gisin, and T. Mor,  Phys. Rev. A  {\bf 51}, 1863 (1995).
\bibitem{s21} M. Koashi and N. Imoto, Phys. Rev. Lett.  {\bf 79}, 2383 (1997).
\bibitem{s22} D. Bru\ss, Phys. Rev. Lett.  {\bf 81}, 3018 (1998).
\bibitem{s23} W. Y. Hwang, I. G. Koh, and Y. D. Han,  Phys.  Lett.  A {\bf 244}, 489 (1998).
\bibitem{s24} A. Cabello,  Phys. Rev. Lett.  {\bf 85}, 5635 (2000).
\bibitem{s25} A. Cabello, Phys. Rev.  A {\bf 61}, 052312 (2000).
\bibitem{s26} G. L. Long and X. S. Liu, Phys. Rev.  A {\bf 65}, 032302 (2002).
\bibitem{s27} P. Xue, C. F. Li, and G. C. Guo, Phys. Rev.  A {\bf 65}, 022317 (2002).
\bibitem{s28} S. J. D. Phoenix, S. M. Barnett, P. D. Townsend, and  K. J. Blow,   J. Modern Optics  {\bf 42}, 1155  (1995).
\bibitem{s29} D. Song,  Phys. Rev. A  {\bf 69}, 034301 (2004).
\bibitem{s30} X. B. Wang,  Phys. Rev. Lett.  {\bf 92}, 077902 (2004).
\bibitem{s31} K. Shimizu and N. Imoto, Phys. Rev. A  {\bf 60}, 157 (1999).
\bibitem{s32} K. Shimizu and N. Imoto, Phys. Rev. A  {\bf 62}, 054303 (2000).
\bibitem{s33} A. Beige {\it et al},   Acta Phys. Pol. A  {\bf 101}, 357 (2002).
\bibitem{s34} A. W$\acute{\rm o}$jcik,  Phys. Rev. Lett.  {\bf 90}, 157901  (2003).
\bibitem{s35} K. Bostr\"{o}m and T. Felbinger,  Phys. Rev. Lett.  {\bf 89}, 187902 (2002).
\bibitem{s36} F. G. Deng, G. L. Long, and X. S. Liu,  Phys. Rev. A   {\bf 68}, 042317 (2003).
\bibitem{s37} F. G. Deng and  G. L. Long,   Phys. Rev. A   {\bf 69}, 052319 (2004).
\bibitem{s38} F. L. Yan and X. Q. Zhang, Euro. Phys. J. B {\bf 41}, 75 (2004).
\bibitem{s39} T. Gao, F. L. Yan, and Z. X. Wang,  Nuovo Cimento B {\bf 119}, 313 (2004).
\bibitem{s40} T. Gao, Z. Naturforsch {\bf 59a}, 597 (2004).
\bibitem{s41} T. Gao, F. L. Yan, and Z. X. Wang, arXiv: quant-ph/0501130.
\bibitem{s42} C. Liu, Z. Dutton, C. H. Behroozi, and  L. V. Hau, Nature  {\bf 409}, 490 (2001).
\bibitem{s43} D. F. Philips {\it et al}, Phys. Rev. Lett. {\bf 86}, 783 (2001).
\bibitem{s44} C. P. Sun, Y. Li, and X. F. Liu, Phys. Rev. Lett.  {\bf 91}, 147903 (2003).
\end{thebibliography}
\end{document}